\documentstyle[11pt]{article}
\input epsf

\newcommand{\beq}{\begin{equation}}
\newcommand{\eeq}{\end{equation}}
\def\smallfrac#1#2{\mbox{\small $\frac{#1}{#2}$}}

\def\smallfrac#1#2{\mbox{\small $\frac{#1}{#2}$}}

\def\punit#1{\hspace{#1\unitlength}}

\def\vert#1#2#3#4{\rule[-2.8\unitlength]{0in}{5.6\unitlength}
\begin{picture}(4,4)(0,-.3)
\put(2,-2.1){\line(0,1){4}}
\put(4.0,0.0){\line(-1,0){4}}
\put(2,-3.2){\makebox(0,0)[b]{\scriptsize \mbox{$#1$}}}
\put(4.4,0){\makebox(0,0)[l]{\scriptsize \mbox{$#2$}}}
\put(2,2.5){\makebox(0,0)[b]{\scriptsize \mbox{$#3$}}}
\put(-.4,0){\makebox(0,0)[r]{\scriptsize \mbox{$#4$}}}
\end{picture}}

\begin{document}

\title{Exact solution and interfacial tension of the six-vertex
model with anti-periodic boundary conditions}

\author{M. T. Batchelor\dag, R. J. Baxter\ddag\S,
M. J. O'Rourke\S\ and C. M. Yung\dag\\
  \dag Department of Mathematics, School of Mathematical Sciences,\\
  \ddag School of Mathematical Sciences, \\
  \S Department of Theoretical Physics, RSPhysSE,\\
Australian National University, Canberra, ACT 0200, Australia\\}

\date{January 1995}

\maketitle

\begin{abstract}
\noindent
We consider the six-vertex model with anti-periodic boundary conditions
across a finite strip. The row-to-row transfer matrix is diagonalised
by the `commuting transfer matrices' method.
{}From the exact solution we obtain an independent derivation of the
interfacial tension of the six-vertex model in the anti-ferroelectric
phase.  The nature of the corresponding integrable
boundary condition on the $XXZ$ spin chain is also discussed.
\end{abstract}

\vfill

\noindent{Short Title:} The anti-periodic six-vertex model

\noindent ANU preprint MRR 012-95, hep-th/9502040

\vfill

\newpage
\section{Introduction and main results}

The six-vertex model and related spin-$\smallfrac{1}{2}$ $XXZ$ chain
play a central role in the
theory of exactly solved lattice models \cite{Baxter82}.
Typically the six-vertex model is `solved' by diagonalising the row-to-row
transfer matrix with periodic boundary conditions.
Several methods have evolved for doing this, including the
co-ordinate Bethe ansatz \cite{Baxter82,LW}, the algebraic Bethe
ansatz \cite{STF,KS}, and the analytic ansatz \cite{R}.
All of these methods rely heavily on the conservation of arrow flux
from row to row of the lattice.

In terms of the vertex weights (see figure 1)
\beq
a = \rho\, \sinh \smallfrac{1}{2} (\lambda-v),
\quad b = \rho\, \sinh \smallfrac{1}{2} (\lambda+v),
\quad c = \rho\, \sinh \lambda
\label{param}
\eeq
the transfer matrix eigenvalues on a strip of width $N$ are
given by \cite{Baxter82}
\beq
\Lambda(v) q(v) = \phi(\lambda-v) q(v+2\lambda')
                + \phi(\lambda+v) q(v-2\lambda')
\label{eigsp}
\eeq
where
\begin{eqnarray}
 \lambda' &=& \lambda - {\rm i} \pi \\
 \phi(v) &=& \rho^N \sinh^N(\smallfrac{1}{2} v) \\
 q(v) &=& \prod_{k=1}^n \sinh \smallfrac{1}{2}(v - v_k).
\end{eqnarray}
The Bethe ansatz equations follow from (\ref{eigsp}) as
\beq
{\phi(\lambda-v_j) \over \phi(\lambda+v_j)} = -
{q(v_j - 2 \lambda') \over q(v_j + 2 \lambda')}, \quad j = 1,\ldots,n.
\eeq
The integer $n$ labels the sectors of the transfer matrix.

Here we consider the same six-vertex model with {\it anti-periodic} boundary
conditions.  That such boundary conditions should preserve integrability
is known through the existence of commuting transfer matrices \cite{dV}.
However, the solution itself has not been found previously.
In section 2 we solve the anti-periodic six-vertex model
by the `commuting transfer matrices'
method \cite{Baxter82}. This approach has its origin in the solution of the
more general 8-vertex model \cite{Baxter72},
which like the present problem, no longer enjoys arrow conservation.
We find the transfer matrix eigenvalues to be given by
\beq
\Lambda(v) q(v) = \phi(\lambda-v) q(v+2\lambda')
                - \phi(\lambda+v) q(v-2\lambda')
\label{eigsap}
\eeq
where now
\beq
q(v) = \prod_{k=1}^N \sinh \smallfrac{1}{4}(v - v_k).
\label{qdef}
\eeq
In this case the Bethe ansatz equations are
\beq
{\phi(\lambda-v_j) \over \phi(\lambda+v_j)} =
{q(v_j - 2 \lambda') \over q(v_j + 2 \lambda')}, \quad j = 1,\ldots,N.
\eeq
In contrast with the periodic case the number of roots is fixed at $N$.

In section 3 we use this solution to derive the interfacial
tension $s$ of the six-vertex model in the anti-ferroelectric regime.
Defining $x = {\rm e}^{-\lambda}$, our final result is
\beq
{\rm e}^{- s/k_B T} = 2 x^{1 \over 2} \prod_{m=1}^{\infty}
\left( {1+x^{4 m} \over 1 + x^{4 m -2} } \right)^2
\eeq
in agreement with the result
obtained from the asymptotic degeneracy of the two largest
eigenvalues \cite{Baxter82,Baxter73}.

With anti-periodic boundary conditions on the vertex model, the
related $XXZ$ Hamiltonian is
\beq
{\cal H} = \sum_{j=1}^{N} \left( \sigma_j^x \sigma_{j+1}^x +
\sigma_j^y \sigma_{j+1}^y + \cosh\!\lambda\,\,
\sigma_j^z \sigma_{j+1}^z \right)
\eeq
where $\sigma^x, \sigma^y$ and $\sigma^z$ are the usual Pauli matrices, with
boundary conditions
\beq
\sigma_{N+1}^x = \sigma_1^x, \quad \sigma_{N+1}^y = -\sigma_1^y, \quad
\sigma_{N+1}^z = -\sigma_1^z.
\eeq
This boundary condition has appeared previously and is amongst
the class of toroidal boundary conditions for which the operator content
of the $XXZ$ chain has been determined by finite-size studies \cite{ABGR}.
It is thus an integrable boundary condition, with the
eigenvalues of the Hamiltonian following from (\ref{eigsap})
in the usual way \cite{Baxter82}, with result
\beq
E = N \cosh\!\lambda\, - \sum_{j=1}^N
{2 \cosh \frac{1}{2} \lambda \, \sinh \lambda \over
 \sinh \frac{1}{2} v_j + \sinh \frac{1}{2} \lambda}.
\eeq

We anticipate that the approach adopted here may also be successful in
solving other models without arrow conservation.
The solution given here can be extended, for example, to the
spin-$S$ generalisation of the six-vertex model/$XXZ$ chain \cite{YB}.

\section{Exact solution}
\setcounter{equation}{0}

To obtain the result (\ref{eigsap}) we adapt where appropriate
the derivation of the periodic result (\ref{eigsp})
(specifically, we refer the reader to Ch. 9 of Ref. \cite{Baxter82}).
We depict a vertex and its corresponding Boltzmann weight graphically, as
\[
\setlength{\unitlength}{.1in}
w(\mu,\alpha | \beta, \mu')  = \punit2
\vert {\alpha}{\mu'}{\beta}{\mu}
\]
where the bond `spins' $\mu,\alpha,\beta, \mu'$ are each
$+1$ if the corresponding arrow points up or to the right and
$-1$ if the arrow points down or to the left.
Thus in terms of the parametrisation (\ref{param}) the nonzero vertex
weights are
\begin{eqnarray}
&w(+,+|+,+) = w(-,-|,-,-) = a& \nonumber\\
&w(+,-|-,+) = w(-,+|,+,-) = b& \\
&w(+,-|+,-) = w(-,+|,-,+) = c& \nonumber
\end{eqnarray}

The row-to-row transfer matrix $T$ has elements
\beq
\setlength{\unitlength}{.1in}
\rule[-2.8\unitlength]{0in}{5.6\unitlength}
T_{\mbox{\boldmath $\alpha \beta$} }
= \sum_{\mu_1} \ldots \sum_{\mu_N}\quad\,\,
\begin{picture}(12,4)(0,-.3)
\put(2,-2.1){\line(0,1){4}}
\put(5,-2.1){\line(0,1){4}}
\put(7,0.0){\line(-1,0){7}}
\put(7.6,0.0){\line(-1,0){.2}}
\put(8.2,0.0){\line(-1,0){.2}}
\put(8.8,0.0){\line(-1,0){.2}}
\put(9.2,0.0){\line(1,0){4}}
\put(11.2,-2.1){\line(0,1){4}}
\put(2,-3.2){\makebox(0,0)[b]{\scriptsize \mbox{$\alpha_1$}}}
\put(2,2.5){\makebox(0,0)[b]{\scriptsize \mbox{$\beta_1$}}}
\put(5,-3.2){\makebox(0,0)[b]{\scriptsize \mbox{$\alpha_2$}}}
\put(5,2.5){\makebox(0,0)[b]{\scriptsize \mbox{$\beta_2$}}}
\put(3,-1.0){\makebox(0,0)[l]{\scriptsize \mbox{$\mu_2$}}}
\put(-.4,0){\makebox(0,0)[r]{\scriptsize \mbox{$\mu_1$}}}
\put(9,-1.0){\makebox(0,0)[l]{\scriptsize \mbox{$\mu_N$}}}
\put(13.6,0){\makebox(0,0)[l]{\scriptsize \mbox{$\mu_{N+1}$}}}
\put(11.2,-3.2){\makebox(0,0)[b]{\scriptsize \mbox{$\alpha_N$}}}
\put(11.2,2.5){\makebox(0,0)[b]{\scriptsize \mbox{$\beta_N$}}}
\end{picture}
\label{tm}
\eeq
where $\mbox{\boldmath $\alpha$} = \{ \alpha_1,\ldots,\alpha_N \}$,
$\mbox{\boldmath $\beta$} = \{ \beta_1,\ldots,\beta_N \}$, and
the anti-periodic boundary condition is such that
$\mu_{N+1}= - \mu_1$.
Now consider an eigenvector $y$ of the form
\beq
y = g_1 \otimes g_2 \otimes \cdots \otimes g_N
\eeq
where $g_i(\alpha_i)$ are two-dimensional vectors.
{}From (\ref{tm}) the product $T y$ can be written as
\beq
\left( T y \right)_{\mbox{\boldmath $\alpha$}}
= {\rm Tr} \left[ G_1(\alpha_1) G_2(\alpha_2)
\cdots G_N(\alpha_N) S \right]
\label{Ty}
\eeq
where $G_i(\pm)$ are $2 \times 2$  matrices with elements
\beq
\setlength{\unitlength}{.1in}
G_i(\alpha)_{\mu \mu'}  = \sum_\beta \punit2
\vert {\alpha}{\mu'}{\beta}{\mu} \qquad g_i(\beta) .
\label{gdef}
\eeq
The appearance of the spin reversal operator
\beq
S = \left( \begin{array}{cc} 0 & 1 \\ 1 & 0 \end{array} \right)
\eeq
in (\ref{Ty}) is the key difference with the periodic case.
However, it does not effect much of the working.
Using (2.1) and (\ref{gdef}) we have
\beq
G_i(+) = \left( \begin{array}{cc}
a\, g_i(+) & 0 \\ c\, g_i(-) & b\, g_i(+) \end{array} \right) \qquad
G_i(-) = \left( \begin{array}{cc}
b\, g_i(-) & c\, g_i(+) \\ 0 & a\, g_i(-) \end{array} \right) .
\label{G}
\eeq
In particular, there still exist the same $2 \times 2$  matrices
$P_1, \ldots , P_N$ such that
\beq
G_i(\alpha) = P_i H_i(\alpha) P_{i+1}^{-1}
\label{GH}
\eeq
where $P_i$ and $H_i$ are of the form
\beq
P_i = \left( \begin{array}{cc}
p_i(+) & \star \\ p_i(-) & \star \end{array} \right) \qquad
H_i(\alpha) = \left( \begin{array}{cc}
g_i'(\alpha) & g_i'''(\alpha) \\ 0 & g_i''(\alpha) \end{array} \right) .
\label{PH}
\eeq

As for the periodic case, (\ref{GH}) follows from the local
`pair-propagation through a vertex' property, i.e. the
existence of $g_i(\alpha), g_i'(\alpha), p_i(\alpha), p_{i+1}(\alpha)$
such that
\beq
\sum_{\beta,\mu'} w(\mu,\alpha | \beta,\mu') g_i(\beta) p_{i+1}(\mu')
= g_i'(\alpha) p_i(\mu)
\eeq
for $\alpha, \mu = \pm 1$. The available parameters are \cite{Baxter82}
\begin{eqnarray}
g_i(+)\!\! &=&\!\! 1, \quad g_i(-) = r_i \,
{\rm e}^{(\lambda+v) \sigma_i/2}  \nonumber\\
g_i'(+)\!\! &=& \!\! a, \quad g_i'(-) = -a\, r_i \,
{\rm e}^{(3\lambda+v) \sigma_i/2} \\
p_i(+)\!\! &=&\!\! 1, \quad p_i(-) = r_i \nonumber
\end{eqnarray}
where $\sigma_i = \pm 1$ and
\beq
r_i = (-)^i r \, {\rm e}^{\lambda(\sigma_1 + \ldots + \sigma_{i-1})}.
\eeq
However, $p_{N+1}$ needs to be different from the periodic case
(where  $p_{N+1}=p_1$). The anti-periodicity suggests that we require
\beq
\left( \begin{array}{c} p_{N+1}(+) \\ p_{N+1}(-) \end{array} \right) =
h \left( \begin{array}{c} p_1(-) \\ p_1(+) \end{array} \right)
\eeq
where $h$ is some scalar. Since we already require $p_i(+) = 1$ and
$p_i(-) = r_i$, we must have
\beq
r_1 = \frac{1}{h} = -r \quad \mbox{and}\quad r_{N+1} = h = \frac{1}{r}.
\eeq
In addition,
\beq
r^2 = (-)^N {\rm e}^{-\lambda(\sigma_1 + \ldots + \sigma_{N})}.
\eeq

To proceed further, we write $P_1$ and $P_{N+1}$ in full,
\beq
P_1 = \left( \begin{array}{cc}
p_1(+) & q_1(+) \\ p_1(-) & q_1(-) \end{array} \right) \qquad
P_{N+1} = \left( \begin{array}{cc}
h\, p_1(-) & q_{N+1}(+) \\ h\, p_1(+) & q_{N+1}(-) \end{array} \right).
\eeq
Then
\begin{eqnarray}
P_{N+1}^{-1} S P_1 &=& \frac{1}{\mbox{det} P_{N+1}}
\left( \begin{array}{cc}
q_{N+1}(-) & - q_{N+1}(+) \\ - h\, p_1(+) & h\, p_1(-) \end{array} \right)
\nonumber\\
&=& \left( \begin{array}{cc}
1/h & \star \\ 0 & -h \frac{\mbox{det} P_1}{\mbox{det} P_{N+1}}
\end{array} \right) .
\end{eqnarray}
Putting the pieces together we then have
\begin{eqnarray}
\left( T y \right)_{\mbox{\boldmath $\alpha$}}
&=& {\rm Tr} \left[ P_1 H_1(\alpha_1)
\cdots H_N(\alpha_N) P_{N+1}^{-1} S \right] \nonumber\\
&=& \frac{1}{h} g_1'(\alpha_1) \cdots g_N'(\alpha_N) -
h \frac{\mbox{det} P_1}{\mbox{det} P_{N+1}}
g_1''(\alpha_1) \cdots g_N''(\alpha_N) .
\end{eqnarray}
However, as for the periodic case, we have
\beq
g_i''(\alpha_i) = a b \frac{g_i^2(\alpha_i) \mbox{det} P_{i+1}}
                           {g_i'(\alpha_i) \mbox{det} P_i}
\eeq
which follows from (\ref{G}) to (\ref{PH}). Thus
\beq
\left( T y \right)_{\mbox{\boldmath $\alpha$}}
= - r g_1'(\alpha_1) \cdots g_N'(\alpha_N)
+ \frac{1}{r} ( a b )^N \frac{g_1^2(\alpha_1) \cdots g_N^2(\alpha_N)}
                             {g_1'(\alpha_1) \cdots g_N'(\alpha_N)}.
\label{Tynew}
\eeq

At this point it is more convenient to write
\beq
y(v) = h_1(v)\otimes h_2(v)\otimes \cdots \otimes h_N(v)
\eeq
where we have defined
\beq
h_i(v) = \left(\begin{array}{c} 1 \\
r_i \, {\rm e}^{\frac{1}{2}(\lambda+v)\sigma_i} \end{array} \right).
\eeq
The result (\ref{Tynew}) can then be more conveniently written as
\beq
T(v) y(v) = - r a^N y(v+2 \lambda') + \frac{1}{r} b^N y(v-2 \lambda').
\label{Tnew}
\eeq
Also let
\beq
y_\sigma^\pm(v) = \exp (-\frac{v}{4} \sum_{i=1}^N \sigma_i ) \,\,
y(\alpha)
\eeq
with $r = \mp \exp ( \frac{\lambda'}{2}\sum_{i=1}^N \sigma_i)$.
Then from (\ref{Tnew}) we have
\beq
T(v) y_\sigma^\pm(v) = \pm \phi(\lambda-v) y_\sigma^\pm(v+2\lambda') \mp
                  \phi(\lambda+v) y_\sigma^\pm(v-2\lambda').
\label{Tmore}
\eeq

To proceed further, let $Q_R^\pm(v)$ be a matrix whose columns are a
linear combination of $y_\sigma^\pm$ with different choices of $\sigma$
($2^N$ altogether). It follows from (\ref{Tmore}) that
\beq
T(v) Q_R^\pm(v) = \pm \phi(\lambda-v) Q_R^\pm(v+2\lambda') \mp
                        \phi(\lambda+v) Q_R^\pm(v-2\lambda') .
\label{TQR}
\eeq
One can show that the transpose of the transfer matrix has the property
$T(-v)=\! ~^t T(v)$. With $Q_L^\mp(v) = \! ~^t Q_R^{\pm}(-v)$
it follows from (\ref{TQR}) that
\beq
Q_L^\pm(v) T(v) = \pm \phi(\lambda-v) Q_L^\pm(v+2\lambda') \mp
 \phi(\lambda+v) Q_L^\pm(v-2\lambda').
\label{TQL}
\eeq
Now let $Q_R(v)=Q_R^+(v)$ and $Q_L(v)=Q_L^+(v)$.\footnote{Equivalent
results are obtained using the other choice of sign.}
Then we can show that the ``commutation relations"
\beq
Q_L(u) Q_R(v) = Q_L(v) Q_R(u)
\label{com}
\eeq
hold for arbitrary $u$ and $v$. This result follows
if we can prove that
$F_{\sigma \sigma'} = \! ~^t y_\sigma^-(-u) y_{\sigma'}^+(v)$ is a symmetric
function of $(u,v)$ for all choices of $\sigma, \sigma'$.
Using (2.24), (2.21) and (2.22) this function reads
\begin{eqnarray}
F_{\sigma \sigma'} &=& \exp \left(\frac{u}{4} \sum_{i=1}^N
\sigma_i - \frac{v}{4} \sum_{i=1}^N \sigma'_i \right)\,
\prod_{j=1}^N \left[ 1 -
(-)^{ \frac{1}{2} \sum_{i=1}^N (\sigma_i + \sigma'_i)} \right. \nonumber\\
&\times & \left.
\exp \left\{ \frac{1}{2}(\lambda-u) \sigma_j +
\frac{1}{2}(\lambda+v) \sigma'_j  - \frac{\lambda}{2}
\left[\sum_{i=j}^N (\sigma_i + \sigma'_i) - \sum_{i=1}^{j-1}
(\sigma_i + \sigma'_i)\right] \right\} \right] .
\end{eqnarray}
Now suppose that in $\sigma, \sigma'$ there are $p$ pairs
$(\sigma_{i_k}, \sigma'_{i_k})$ where
$\sigma_{i_k} + \sigma'_{i_k} = 0$ with $k=1,\ldots,p$.
The terms in $F_{\sigma \sigma'}$ which involve these $\sigma_{i_k}$
(in the prefactor and in the $j=i_k$ terms) are manifestly symmetric
in $(u,v)$. The remaining terms are exactly of the form (2.29) with
$N \rightarrow N-p$ after relabelling of sites. We can thus restrict ourselves
to the case where $\sigma_i = \sigma'_i, i = 1,\ldots,N'$, for all $N'$.
To prove this case we proceed inductively. From (2.28) we have
\beq
F_{\sigma \sigma} = \prod_{j=1}^N \left[ {\rm e}^{
\frac{1}{4}(u-v) \sigma_j} - (-)^N {\rm e}^{
\frac{1}{4}(v-u) \sigma_j }\, {\rm e}^{ -\lambda (
\sigma_N + \ldots + \sigma_{j+1})} \, {\rm e}^{\lambda (
\sigma_{j-1} + \ldots + \sigma_1)}
\right] .
\label{Feq}
\eeq
Let us now denote $F_{\sigma \sigma} = F_N(\sigma_1,\ldots,\sigma_N)$.
By inspection, $F_1(\sigma_1)$ and $F_2(\sigma_1,\sigma_2)$ are
symmetric in $(u,v)$. Suppose
$F_{N-2}(\sigma_1,\ldots,\sigma_{N-2})$ is symmetric in $(u,v)$
and furthermore that $\sigma_k + \sigma_{k+1} = 0$ for some $k$.
Then from (\ref{Feq}) we have
$F_N(\sigma_1,\ldots,\sigma_k,-\sigma_k,\ldots,\sigma_N) =
F_{N-2}(\sigma_1,\ldots,\hat{\sigma}_k,\hat{\sigma}_{k+1},
\ldots,\sigma_{N-2})$
times a symmetric function of $(u,v)$, which is therefore symmetric
in $(u,v)$. This is true for all $1 \le k \le N-1$. The only case left to
consider is therefore $\sigma_1=\sigma_2=\ldots=\sigma_N$.
But from (\ref{Feq}) we have $F_N(\sigma_1,\sigma_2,\ldots,\sigma_{N-1},
\sigma_1) = F_{N-2}(\sigma_2,\ldots,\sigma_{N-1})$ times a symmetric
function of $(u,v)$, which is again symmetric.
Thus by induction on $N$, the assertion (\ref{com}) follows.

As in the periodic case, we assume that $Q_R(v)$ is invertible at
some point $v=v_0$ and define
\beq
Q(v) = Q_R(v) Q_R^{-1}(v_0) = Q_L^{-1}(v_0) Q_L(v).
\eeq
Then from (\ref{TQL}) and (\ref{com}) we obtain
\beq
T(v) Q(v) = Q(v) T(v) = \phi(\lambda-v) Q(v+2\lambda') -
                        \phi(\lambda+v) Q(v-2\lambda')
\label{TQ}
\eeq
and $Q(u) Q(v) = Q(v) Q(u)$.
This allows $T(v), Q(v)$ and $Q(v\pm 2 \lambda')$ to be
simultaneously diagonalised, yielding the relation
(\ref{eigsap}) for their eigenvalues.
The precise functional form of the eigenvalue $q(v)$ of $Q(v)$,
given in (\ref{qdef}), follows from
(\ref{TQ}) by noting that $T(v+2\pi {\rm i}) = - T(v)$ and
considering the limits $v\rightarrow\pm\infty$.

\section{Interfacial tension}
\setcounter{equation}{0}

In this section we derive the interfacial tension by solving
the functional relation (\ref{eigsap}) and
integrating over the band of largest
eigenvalues of the transfer matrix \cite{JKM}.
We consider the case where $N$, the number of columns in the
lattice, is even.
The partition function of the model is expressed in terms of the
eigenvalues $\Lambda(v)$ of the row-to-row transfer matrix $T(v)$ as
\beq
Z = \sum \left[ \Lambda (v) \right]^M
\label{partfundef}
\eeq
where the sum is over all $2^N$ eigenvalues.

The interfacial tension is defined as follows.
Consider a single row of the lattice.
For a system with periodic boundary conditions, in the
$\lambda \rightarrow \infty$ limit we see from (\ref{param})
that the vertex weight $c$ is much greater than
the weights $a$ and $b$,
so in this limit, the row can be in one of two possible
anti-ferroelectrically ordered ground states.
These are made up entirely of spins
with Boltzmann weight $c$, and are related to one another by arrow-reversal.

When we impose anti-periodic boundary conditions, this ground-state
configuration is no longer consistent with $N$ even.
To ensure the anti-periodic boundary condition, vertices with Boltzmann
weight $c$ must occur an odd number of times in each row.
Thus the lowest-energy configuration for the row
in the $\lambda \rightarrow \infty$ limit will consist of $N-1$
vertices with weight $c$, and one vertex of either types $a$ or $b$.
This different vertex can occur anywhere in the row.

As we add rows to form the lattice, the $a$ or $b$ vertex
in each row forms a ``seam'' running approximately vertically
down the lattice; it can jump from left to right but the mean direction
is downwards.\footnote{
This is the analogue of the anti-ferromagnetic seam in the
Ising model \cite{onsager}.}
A typical lowest-energy configuration is shown in figure 2.
The extra free energy due to this seam is called the
interfacial tension.
This will grow with the height $M$ of the lattice, so we expect
that for large $N$ and $M$
the partition function of the lattice will be of the form
\beq
Z \sim \exp\left[ ( -N\!M\! f - M\!s)/k_BT\right]
\label{partfun}
\eeq
where $f$ is the normal bulk free energy, and $s$ is the interfacial tension.

We introduce the variables
\beq
x = e^{-\lambda}, \;\;\;\;z=e^{-v/2}.
\eeq
Expressing the Boltzmann weights in terms of $z$ and $x$, from
(\ref{param}) the model is physical when $z$ and $x$ are real, and $z$
lies in the interval
\beq
x^{1/2} \leq z \leq x^{-1/2}.
\label{interval}
\eeq
We consider $\lambda \geq 0$ in order that the Boltzmann weights are
non-negative, so we must have $x \leq 1$.
Let
\beq
\tilde{Q}(z) = \prod_{j=1}^{N} (z-z_{j})
\label{thisone}
\eeq
where $z_{j} = e^{-v_{j}/2}$, $j=1,\ldots,N$, and
\beq
V(z) = \Lambda(v) (2z\rho^{-1})^{N} (-)^{N/2}.
\label{vandlambda}
\eeq
In terms of these variables, the functional relation (\ref{eigsap})
becomes
\beq
\tilde{Q}(z) V(z) = (1-z^{2}x^{-1})^{N} \tilde{Q}(-zx) -
(1-z^{2}x)^{N} \tilde{Q}(-zx^{-1}).
\label{eigsap2}
\eeq
Both terms on the right hand side of (\ref{eigsap2}) are polynomials in $z$ of
degree $3N$, but the coefficients of 1 and $z^{3N}$ vanish,
so $z^{-1}V(z)$ is a polynomial in $z$ of degree $2N-2$.
We know how to solve equations of this form
for both $V(z)$ and $\tilde{Q}(z)$ using Wiener-Hopf
factorisations (see references \cite{Baxter72,Baxter73} and \cite{noble}).

We shall need some idea where the zeros of the
polynomials $\tilde{Q}(z)$ and $V(z)$ lie in order to construct the Wiener-Hopf
factorisations.
{}From the anti-periodicity of $T(v)$ we see that
$V(z)$ is an odd function of $z$,
\beq
V(-z) = -V(z)
\eeq
so its zeros and poles must occur in plus--minus pairs.
To locate the zeros in the $z$-plane,
we consider $z$ to be a free variable, and vary the parameter $x$, in
particular looking at the limit $x \rightarrow 0$.

We find the following; in the $x \rightarrow 0$ limit,
$N-2$ of the $N$ zeros of $\tilde{Q}(z)$ lie on
the unit circle, the other two lying at distances proportional to
$x^{1/2}$ and $x^{-1/2}$.
For $V(z)$, there is the simple zero at the origin, and two zeros on the
unit circle.
The remaining $2N-4$ zeros of $V(z)$ are divided into two sets, with
$N-2$ of them that approach the origin and $N-2$
that approach $\infty$ as $x \rightarrow 0$.
The $N$ zeros of the two polynomials that lie on the unit circle are
spaced evenly around the circle.

As $x$ is increased, the zeros of $\tilde{Q}(z)$ and
$z^{-1}V(z)$ will all shift.
We assume that the distribution
of the zeros mentioned above does not change significantly as $x$
increases.
Thus the zeros that lie at the origin in the $x \rightarrow 0$ limit
move out from the origin as $x$ increases, but not so far out as the
unit circle, and similarly for the zeros that lie at $\infty$.
Also, the zeros that lie on the unit circle are assumed to stay in some
neighbourhood of the unit circle as $x$ increases
(we will show that these zeros remain
exactly on the unit circle as $x$ increases,
which is what happens in the periodic boundary condition case).

Bearing in mind the above comments, we write
\beq
\tilde{Q}(z) = \tilde{Q}_{1}(z) (z-\alpha)(z-\beta^{-1})
\eeq
where $\tilde{Q}_1(z)$ is a polynomial of degree $N-2$ whose zeros
are $O(1)$ as $x \rightarrow 0$, and $\alpha, \beta = O(x^{1/2})$, so
$\alpha$ lies inside the unit circle, $\beta^{-1}$ outside.

Also, let $V(z) = z(z-t_{1})(z-t_{2}) A(z) B(z)$,
where $A(z)$ and $B(z)$ are both polynomials of degree $N-2$,
the zeros of $A(z)$ being all the zeros of $V(z)$ that lie inside the
unit circle, $B(z)$ containing all those that lie outside, and $t_1$ and
$t_2$ are the zeros that lie on the unit circle.
Since $V(z)$ is an odd function,
both $A(z)$ and $B(z)$ must be even functions of $z$,
and we must have $t_{1} = -t_{2}$, so
letting $t_{1}=-t_{2}=  t$, we write
\beq
V(z) = z(z^{2}-t^{2})A(z) B(z).
\label{vfactored}
\eeq

Draw the contours ${\cal C}_{+}$ and ${\cal C}_{-}$ in the complex
$z$-plane, both oriented in the positive direction,
with ${\cal C}_{-}$ outside the unit
circle, ${\cal C}_{+}$ outside ${\cal C}_{-}$, and such that
there are no zeros
of $\tilde{Q}(z)$ or $V(z)$ on the boundary of or inside the annulus between
${\cal C}_{-}$ and ${\cal C}_{+}$.
Then $\beta^{-1}$ and all the zeros of $B(z)$ lie
outside ${\cal C}_+$ (see figure 3).

Define $r(z)$ as the quotient of the two terms in the RHS of the
functional relation (\ref{eigsap2});
\beq
r(z) = -\frac{ \tilde{Q}(-zx^{-1})(1-z^{2}x)^{N}}
{\tilde{Q}(-zx) (1-z^{2}x^{-1})^{N}}
\eeq
($r(z)$ has no zeros or poles on or between the curves ${\cal C}_+$ and
${\cal C}_-$).
Then in the $x \rightarrow 0$ limit, we see that
$|r(z)| \sim 1/z^N$, so when $|z| > 1$, $|r(z)|<1$.
Thus $\ln[1+r(z)]$ can be chosen to be single-valued and analytic when
$z$ lies in the annulus between ${\cal C}_-$ and ${\cal C}_+$.
We can therefore make a Wiener-Hopf factorisation of $1+r(w)$ by
defining the functions $P_{+}(z)$ and $P_{-}(z)$ as
\beq
\ln P_{\pm}(z) = \pm \frac{1}{2\pi i} \oint _{{\cal C}_{\pm}}
\ln \left[1+r(z') \right] \frac{dz'}{z'-z}
\eeq
Then $P_{+}(z)$ is an analytic and non-zero (ANZ) function of $z$
for $z$ inside ${\cal C}_{+}$, and $P_{-}(z)$ is an ANZ function of
$z$ for $z$ outside ${\cal C}_{-}$.
As $|z| \rightarrow \infty$, we note that $P_{-}(z) \rightarrow 1$.
When $z$ is inside the annulus between ${\cal C}_-$ and ${\cal C}_+$,
we have, by Cauchy's integral formula
\beq
1+r(z) = P_{+}(z) \: P_{-}(z)
= \frac{V(z) \tilde{Q}(z)}{\tilde{Q}(-zx)(1-z^{2}x^{-1})^{N}}.
\label{analytcont}
\eeq
%When $z$ is outside ${\cal C}$, (\ref{analytcont}) defines the analytic
%continuation of either $P_{+}(z)$ or $P_{-}(z)$.
We then define the functions $V_{\pm}(z)$;
\begin{eqnarray}
V_{+}(z) & = & P_{+}(z) \tilde{Q}(-zx)/(z-\beta^{-1})  \label{vp} \\
V_{-}(z) & = & P_{-}(z) (1-z^{2}x^{-1})^{N} \left/ \left[
\tilde{Q}_{1}(z)(z-\alpha)
\right.
\label{vm}
\right]
\end{eqnarray}
where $V_{+}(z)$ is an ANZ function of $z$ for $z$
inside ${\cal C}_{+}$, $V_{-}(z)$ an ANZ function of $z$
for $z$ outside ${\cal C}_{-}$.
We have split $V(z)$ into two
factors, $V_+(z)$ and $V_-(z)$, with $V(z) = V_+(z) V_-(z)$ when $z$ is
between ${\cal C}_+$ and ${\cal C}_-$.

Equating (\ref{vfactored}) with the expression $V(z) = V_{+}(z)V_{-}(z)$
we have
\beq
\frac{V_{+}(z) }{B(z)} = \frac{A(z)}{V_{-}(z)} z(z^{2}-t^2).
\eeq
The LHS (RHS) is an ANZ function of $z$ inside ${\cal C}_{+}$ (outside
${\cal C}_{-}$), which is bounded as $|z| \rightarrow \infty$
and so the function must be a constant, $c_{1}$ say.
Thus
\begin{eqnarray}
V_{+}(z) & = & c_{1} B(z)                 \label{v1} \\
V_{-}(z) & = & c_{1}^{-1}z(z^{2}-t^2)A(z). \label{v2}
\end{eqnarray}

When $|z|<1$, we proceed the same way.
Draw the curves ${\cal C}_{+}'$ and ${\cal C}_{-}'$,
${\cal C}'_{+}$ inside the unit circle,
${\cal C}_{-}'$ inside ${\cal C}_{+}'$, and with $\alpha$ and all the zeros of
$A(z)$ inside ${\cal C}_{-}'$.

In the limit $x\rightarrow 0$, $|1/r(z)| \sim z^N$, so $|1/r(z)|<1$.
Thus $\ln [1+1/r(z)]$ can be chosen to be single-valued and analytic
between and on ${\cal C}_{+}'$ and ${\cal C}_{-}'$.
We can then Wiener-Hopf factorise $1+1/r(z)$ by defining
the functions $P_{+}'(z)$ and $P_-'(z)$ as
\beq
\ln P_{\pm}'(z) = \pm \frac{1}{2\pi i} \oint_{{\cal C}'_{\pm}} \ln
\left[ 1+\frac{1}{r(z')} \right] \frac{dz'}{z'-z},
\eeq
where $P_{+}'(z)$ is ANZ inside ${\cal C}_{+}'$, $P_{-}'(z)$ is ANZ for $z$
outside ${\cal C}_{-}'$.
As $|z| \rightarrow \infty$, $P_{-}'(z) \rightarrow 1$.
When $z$ is in the annulus between ${\cal C}_+'$ and ${\cal C}_-'$,
Cauchy's integral formula now implies
\beq
1 + \frac{1}{r(z)} = P_{+}'(z) \: P_{-}'(z) =
-\frac{V(z)\tilde{Q}(z)}{\tilde{Q}(-zx^{-1})(1-z^{2}x)^{N}}.
\eeq
Define $V'_{+}(z)$ and $V_-'(z)$ as follows:
\begin{eqnarray}
V'_{+}(z) & = & P'_{+}(z) (1-z^{2}x)^{N} \left/
\left[\tilde{Q}_{1}(z)(z-\beta^{-1})\right] \right.
\label{vpp} \\
V_{-}'(z) & = & P'_{-}(z) \tilde{Q}(-zx^{-1})/(z-\alpha). \label{vmp}
\end{eqnarray}
We have now factorised $V(z)$ into two factors,
$V_{+}'(z)$ which is is ANZ
for $z$ inside ${\cal C}'_{+}$, and $V'_{-}(z)$ which is ANZ for $z$ outside
${\cal C}'_{-}$.
When $z$ is in the annulus between ${\cal C}_+'$ and ${\cal C}_-'$,
we have the equality $V(z) = V_+'(z) V_-'(z)$.

When $z$ is inside this annulus, we equate (\ref{vfactored}) with $V(z)=
V'_{+}(z) V'_{-}(z)$ to get
\beq
\frac{V_{+}'(z)}{B(z)(z^2-t^2)} = \frac{zA(z)}{V_{-}'(z)}
\eeq
where now the LHS (RHS) is an ANZ function of $z$ for $z$ inside ${\cal
C}'_{+}$ (outside ${\cal C}'_{-}$).
Thus both sides of the equation are constant, $c_2$ say, and we have
\begin{eqnarray}
V'_{+}(z) & = & c_{2} (z^2-t^2)B(z) \label{v3} \\
V'_{-}(z) & = & c_{2}^{-1}zA(z).    \label{v4}
\end{eqnarray}
{}From equations (\ref{v1}), (\ref{v3}) and (\ref{v2}),
(\ref{v4}), we have the following
\begin{eqnarray}
V_{+}'(z) & = & (c_{2}/c_{1}) V_{+}(z) (z^2-t^2)  \label{vppvp}\\
V_{-}(z) & = & (c_{1}/c_{2}) V_{-}'(z) (z^2-t^2). \label{vmvmp}
\end{eqnarray}
To evaluate the constant $c_1/c_2$, consider (\ref{vmvmp}) in the limit $z
\rightarrow \infty$; we noted earlier that $P_-(z)$, $P_-'(z)
\rightarrow 1$ as $z \rightarrow \infty$, so from (\ref{thisone}),
(\ref{vm}) and (\ref{vmp}) we deduce that
\beq
c_1/c_2 = 1.
\eeq

We may use equations (\ref{vppvp}) and (\ref{vmvmp}) to derive
recurrence relations satisfied by $\tilde{Q}(z)$, which we can solve explicitly
in the $N \rightarrow \infty$ limit.

{}From equations (\ref{vp}), (\ref{vpp}) and (\ref{vppvp}),
we deduce the recurrence relation
\beq
\tilde{Q}(z) \: \tilde{Q}(-z x) = (1-z^{2}x)^{N}
\frac{(z-\alpha)(z-\beta^{-1})}{(z^2-t^2)} \frac{P_{+}'(z)}{P_{+}(z)}
\label{recur}
\eeq
valid when $z$ is inside ${\cal C}_+'$.
In the limit $N \rightarrow \infty$, the $P_{+}$ and $P_{+}'$
functions $\rightarrow 1$, so we find that $\tilde{Q}(z)$ is given by
\beq
\tilde{Q}(z)  =  \tilde{Q}(0) \prod_{m=1}^{\infty}
\left( \frac{1-z^{2} x^{4m-3}}{1-z^{2}x^{4m-1}} \right)^{N} \!
\frac{(1-z^2 t^{-2} x^{4m-2})}{(1-z^2 t^{-2}x^{4m-4})}
\frac{(1-z \alpha^{-1} x^{2m-2})}{(1+z\alpha^{-1}x^{2m-1})}
\frac{(1-z\beta x^{2m-2})}{(1+z\beta x^{2m-1})} .
\label{wl1}
\eeq
This still contains the parameters $t$, $\alpha$ and $\beta$.
{}From (\ref{recur}) in the $N \rightarrow \infty$ limit,
setting $z=0$ we note that
\beq
\left[ \tilde{Q}(0) \right]^2 = -t^{-2} \alpha\beta^{-1}  .
\eeq

{}From equations (\ref{vm}), (\ref{vmp}) and
(\ref{vmvmp}), we get the recurrence relation
\beq
\tilde{Q}(z) \: \tilde{Q}(-z x^{-1}) = (1-z^{2}x^{-1})^{N}
\frac{(z-\alpha)(z-\beta^{-1})}{(z^2-t^2)} \frac{P_{-}(z)}{P_{-}'(z)}
\eeq
which is valid for $z$ outside ${\cal C}_-$.
Taking the limit $N \rightarrow \infty$ once more, so that the
functions $P_{-}(z)$ and $P_{-}'(z) \rightarrow 1$, we get
\beq
\tilde{Q}(z)  =  z^{N} \prod_{m=1}^{\infty}
\left( \frac{1-z^{-2} x^{4m-3}}{1-z^{-2}x^{4m-1}} \right)^{N}
\frac{(1-z^{-2} t^2 x^{4m-2})}{(1-z^{-2}t^2 x^{4m-4})}
\frac{(1-z^{-1}\alpha  x^{2m-2})}{(1+z^{-1}\alpha x^{2m-1})}
\frac{(1-z^{-1}\beta^{-1} x^{2m-2})}{(1+z^{-1} \beta^{-1} x^{2m-1})}.
\label{wg1}
\eeq

To derive an expression for $V(z)$ valid between ${\cal C}_+$ and ${\cal
C}_-'$, using equation (\ref{vmvmp}), we have
\begin{eqnarray}
V(z) & = & V_{+}(z) V_{-}'(z) (z^2-t^2)  \nonumber\\
     & = & \tilde{Q}(-zx) \tilde{Q}(-zx^{-1})(z^2-t^2) \left/
           \left[ (z-\alpha)(z-\beta^{-1})\right] \right. . \label{v}
\end{eqnarray}
We use (\ref{wl1}) for $\tilde{Q}(-zx)$ and (\ref{wg1}) for
$\tilde{Q}(-zx^{-1})$, and substitute into equation (\ref{v}).
This produces a lengthy expression for $V(z)$ involving the parameters
$\alpha, \beta$ and $t$, which simplifies when one considers the
oddness of $V(z)$.
The poles of $V(z)$ must occur in pairs, and this is only possible if
$\alpha$ and $\beta$ are related by
\beq
\alpha \beta = - x.
\eeq
Substituting this in, the infinite products involving $\alpha$ and
$\beta$ cancel, and we get, from (\ref{vandlambda}) and (\ref{v})
\beq
\Lambda(v) = G(z/t) \: (\rho/2x)^N
\prod_{m=1}^{\infty}  \left(
           \frac{1-z^{2}x^{4m-1}}{1-z^{2}x^{4m+1}}\cdot
           \frac{1-z^{-2}x^{4m-1}}{1-z^{-2}x^{4m+1}}  \right)^{N}
\label{maxeig}
\eeq
where
\beq
G(z) = \pm i x^{1/2} (z-z^{-1}) \prod_{m=1}^{\infty}
\left( \frac{1-z^2 x^{4m}}{1-z^{2}x^{4m-2}} \cdot
\frac{1-z^{-2}x^{4m}}{1-z^{-2}x^{4m-2}} \right)  .
\label{GG}
\eeq
This expression for the eigenvalue is still dependent on the parameter
$t$, different values of $t$ corresponding to different
eigenvalues of the transfer matrix.
All we know about $t$ so far is that it is bounded as $x \rightarrow 0$,
and that it lies on the unit circle in the $x \rightarrow 0$ limit.
We shall now show that it in fact remains exactly on the unit circle
as $x$ increases.

We substitute into the functional relation (\ref{eigsap2}),
using equations (\ref{wl1}) and (\ref{wg1})
to get an expression for the product $\tilde{Q}(z)V(z)$ which is valid when
$z$ is in the annulus between ${\cal C}_+$ and ${\cal C}_-'$.
%We can use this expression both to verify the assertion that the $N-2$
%zeros of $\tilde{Q}(z)$ lie on the unit
%circle in the limit $x \rightarrow 0$ remain there as we increase $x$,
%and to show that $t$ remains on the unit circle also.
Substituting into (\ref{eigsap2}), the function on the right hand side
is equal to zero when $z$ is one of the $N-2$ zeros of $\tilde{Q}_1(z)$,
or when $z = \pm t$.
For the latter case, substituting $z = t$ and $-t$, and dividing the
resulting equations, we arrive at the following relation between
$\alpha, x, $ and $t$
\beq
\alpha^2 = -t^2 x
\label{alph}
\eeq
which means that $t$ must satisfy
\beq
\left[ \phi(t) \right]^N = \pm 1
\label{phi}
\eeq
where $\phi(t)$ is given by
\beq
\phi(t) = t \prod_{m=1}^{\infty} \left(
\frac{1-t^2 x^{4m-1}}{1-t^2 x^{4m-3}} \cdot
\frac{1-t^{-2} x^{4m-3}}{1-t^{-2}x^{4m-1}} \right).
\eeq
This implies that $t$ lies on the unit circle for all $x$,
there being $2N$ possible choices for $t$.
The partition function depends on $t$ only via $t^2$, so there are
only $N$ distinct eigenvalues.
The right hand side of (\ref{eigsap2}) also vanishes when $z$ is a zero of
$\tilde{Q}_1(z)$ so in the same way we show that
the zeros of $\tilde{Q}_1(z)$ lie exactly on the unit circle
for all $x$.
As the zeros lie exactly on the unit circle, we may shift the curves
${\cal C}_-$ and ${\cal C}_+'$ so that they just surround the unit
circle.
Hence our expressions for $\tilde Q(z)$ are valid all the way up to the unit
circle;
(\ref{wl1}) is valid for $|z|<1$, and (\ref{wg1}) is valid for $|z|>1$.

We now evaluate the partition function, as defined in
(\ref{partfundef}), in the large-lattice limit.
When $v$ is real, the eigenvalues (\ref{maxeig}) are complex, so as $N
\rightarrow \infty$, the
partition function, a sum over the $N$ eigenvalues defined by
(\ref{phi}), becomes an integral over all the allowed values of $t$,
\beq
Z = \oint \rho(t) \left[ \Lambda (v) \right]^M dt
\label{pfint}
\eeq
where the integral is taken around the unit circle,
and $\rho(t)$ is some distribution function, independent of $N$ and $M$.
Substituting (\ref{v}) into (\ref{pfint}) then gives an expression for
$Z$.
(The number of rows $M$ is even to ensure periodic boundary conditions
vertically, and so the $\pm$ sign in (\ref{GG}) is irrelevant.)

The eigenvalue (\ref{maxeig})
contains two distinct types of factors; those that are
powers of $N$, and those that are not.
The terms that increase exponentially with $N$ contribute to the bulk part of
the partition function, the free energy per site
in the thermodynamic limit.
This factor is also independent of $t$, and can be taken out of
the integral (\ref{pfint}).
The integral is then independent of $N$, so
we have, from (\ref{partfun})
\beq
e^{-f/k_BT} = (\rho/2x) \prod_{m=1}^{\infty}
\left( \frac{1-z^{2}x^{4m-1}}{1-z^2 x^{4m+1}} \cdot
\frac{1-z^{-2}x^{4m-1}}{1-z^{-2}x^{4m+1}} \right)
\eeq
for the free energy per site in the thermodynamic limit.
This result agrees with the result for periodic
boundary conditions (equations (8.9.9) and (8.9.10) of
Ref. \cite{Baxter82}).

{}From equation (\ref{partfun}), the other factors in (\ref{v}) make up the
interfacial tension, given by
\beq
e^{-Ms/k_BT} = \oint \rho(t) \left[ G(z/t) \right] ^M dt .
\label{intten}
\eeq
For $M$ sufficiently large, we may evaluate this integral using
saddle-point integration.
The integral is given by the value of the integrand at its saddle point,
together with some multiplicative factors that we can disregard as $M
\rightarrow \infty$.
The function $G$ satisfies the relation
\beq
G(z) = G(-1/z)
\eeq
which implies that the function has a saddle point when $z = \pm i$.
Hence the integrand in (\ref{intten}) is maximised when
\beq
t = t_{\rm saddle} = \pm i z.
\eeq
As $z$ is arbitrary, this point may lie off the unit circle; it will
however lie inside the annulus between ${\cal C}_+$ and ${\cal C}'_-$
because of the restriction (\ref{interval}), and so we will be able to
deform the contour to pass through this saddle point.
Hence we arrive at the final result
\beq
\mbox{\rm e}^{-s/k_B T} = 2 x^{1/2} \prod_{m=1}^{\infty} \left(
\frac{1+x^{4m}}{1+x^{4m-2}} \right)^2.
\eeq

\section*{Acknowledgements}

MTB and CMY thank the Australian Research Council for financial support.

\newpage
\begin{figure}
\setlength{\unitlength}{0.012500in}%
\begingroup\makeatletter
% extract first six characters in \fmtname
\def\x#1#2#3#4#5#6#7\relax{\def\x{#1#2#3#4#5#6}}%
\expandafter\x\fmtname xxxxxx\relax \def\y{splain}%
\ifx\x\y   % LaTeX or SliTeX?
\gdef\SetFigFont#1#2#3{%
  \ifnum #1<17\tiny\else \ifnum #1<20\small\else
  \ifnum #1<24\normalsize\else \ifnum #1<29\large\else
  \ifnum #1<34\Large\else \ifnum #1<41\LARGE\else
     \huge\fi\fi\fi\fi\fi\fi
  \csname #3\endcsname}%
\else
\gdef\SetFigFont#1#2#3{\begingroup
  \count@#1\relax \ifnum 25<\count@\count@25\fi
  \def\x{\endgroup\@setsize\SetFigFont{#2pt}}%
  \expandafter\x
    \csname \romannumeral\the\count@ pt\expandafter\endcsname
    \csname @\romannumeral\the\count@ pt\endcsname
  \csname #3\endcsname}%
\fi
\endgroup
%\begin{picture}(460,82)(60,649)
\begin{picture}(460,82)(60,800)
\thicklines
\put( 60,700){\vector( 1, 0){ 60}}
\put( 60,700){\vector( 1, 0){ 10}}
\put( 90,670){\vector( 0, 1){ 10}}
\put(200,700){\vector(-1, 0){ 60}}
\put(200,700){\vector(-1, 0){ 10}}
\put( 90,670){\vector( 0, 1){ 60}}
\put(170,731){\vector( 0,-1){ 60}}
\put(170,730){\vector( 0,-1){ 10}}
\put(220,700){\vector( 1, 0){ 60}}
\put(362,700){\vector(-1, 0){ 60}}
\put(250,730){\vector( 0,-1){ 60}}
\put(250,730){\vector( 0,-1){ 10}}
\put(220,700){\vector( 1, 0){ 10}}
\put(360,700){\vector(-1, 0){ 10}}
\put(330,670){\vector( 0, 1){ 60}}
\put(330,670){\vector( 0, 1){ 10}}
\put(410,670){\vector( 0, 1){ 60}}
\put(380,700){\line( 1, 0){ 60}}
\put(490,730){\line( 0,-1){ 60}}
\put(460,700){\vector( 1, 0){ 60}}
\put(490,670){\vector( 0, 1){ 10}}
\put(490,730){\vector( 0,-1){ 10}}
\put(470,700){\vector(-1, 0){ 10}}
\put(440,700){\vector(-1, 0){ 10}}
\put(410,680){\vector( 0,-1){ 10}}
\put(380,700){\vector( 1, 0){ 10}}
\put( 86,650){\makebox(0,0)[lb]{\smash{\SetFigFont{17}{20.4}{it}a}}}
\put(166,650){\makebox(0,0)[lb]{\smash{\SetFigFont{17}{20.4}{it}a}}}
\put(246,649){\makebox(0,0)[lb]{\smash{\SetFigFont{17}{20.4}{it}b}}}
\put(326,649){\makebox(0,0)[lb]{\smash{\SetFigFont{17}{20.4}{it}b}}}
\put(407,651){\makebox(0,0)[lb]{\smash{\SetFigFont{17}{20.4}{it}c}}}
\put(486,650){\makebox(0,0)[lb]{\smash{\SetFigFont{17}{20.4}{it}c}}}
\end{picture}
\label{fig:vertices}
\end{figure}
\begin{center}
Figure 1: Standard vertex configurations and corresponding weights.
\end{center}

\begin{figure}
\setlength{\unitlength}{.4mm}
\begin{picture}(200,120)(-10,275)
\put(40,10) {\line(0,1){180}}
\put(80,10) {\line(0,1){180}}
\put(120,10) {\line(0,1){180}}
\put(160,10) {\line(0,1){180}}
\put(200,10) {\line(0,1){180}}
\put(240,10) {\line(0,1){180}}
\put(280,10) {\line(0,1){180}}
\put(320,10) {\line(0,1){180}}
\put(10,40) {\line(1,0){340}}
\put(10,80) {\line(1,0){340}}
\put(10,120) {\line(1,0){340}}
\put(10,160) {\line(1,0){340}}
\thicklines
\put(40,30) {\vector(0,1){0}}
\put(40,50) {\vector(0,-1){0}}
\put(40,110) {\vector(0,1){0}}
\put(40,130) {\vector(0,-1){0}}
\put(40,190) {\vector(0,1){0}}
\put(80,10) {\vector(0,-1){0}}
\put(80,70) {\vector(0,1){0}}
\put(80,90) {\vector(0,-1){0}}
\put(80,150) {\vector(0,1){0}}
\put(80,170) {\vector(0,-1){0}}
\put(120,30) {\vector(0,1){0}}
\put(120,70) {\vector(0,1){0}}
\put(120,110) {\vector(0,1){0}}
\put(120,130) {\vector(0,-1){0}}
\put(120,190) {\vector(0,1){0}}
\put(160,30) {\vector(0,1){0}}
\put(160,50) {\vector(0,-1){0}}
\put(160,110) {\vector(0,1){0}}
\put(160,150) {\vector(0,1){0}}
\put(160,190) {\vector(0,1){0}}
\put(200,10) {\vector(0,-1){0}}
\put(200,70) {\vector(0,1){0}}
\put(200,90) {\vector(0,-1){0}}
\put(200,150) {\vector(0,1){0}}
\put(200,170) {\vector(0,-1){0}}
\put(240,30) {\vector(0,1){0}}
\put(240,50) {\vector(0,-1){0}}
\put(240,110) {\vector(0,1){0}}
\put(240,130) {\vector(0,-1){0}}
\put(240,190) {\vector(0,1){0}}
\put(280,10) {\vector(0,-1){0}}
\put(280,70) {\vector(0,1){0}}
\put(280,90) {\vector(0,-1){0}}
\put(280,150) {\vector(0,1){0}}
\put(280,170) {\vector(0,-1){0}}
\put(320,30) {\vector(0,1){0}}
\put(320,50) {\vector(0,-1){0}}
\put(320,110) {\vector(0,1){0}}
\put(320,130) {\vector(0,-1){0}}
\put(320,190) {\vector(0,1){0}}
\put(10,40) {\vector(-1,0){0}}
\put(30,80) {\vector(1,0){0}}
\put(10,120) {\vector(-1,0){0}}
\put(30,160) {\vector(1,0){0}}
\put(70,40) {\vector(1,0){0}}
\put(50,80) {\vector(-1,0){0}}
\put(70,120) {\vector(1,0){0}}
\put(50,160) {\vector(-1,0){0}}
\put(90,40) {\vector(-1,0){0}}
\put(110,80) {\vector(1,0){0}}
\put(90,120) {\vector(-1,0){0}}
\put(110,160) {\vector(1,0){0}}
\put(130,40) {\vector(-1,0){0}}
\put(150,80) {\vector(1,0){0}}
\put(150,120) {\vector(1,0){0}}
\put(130,160) {\vector(-1,0){0}}
\put(190,40) {\vector(1,0){0}}
\put(170,80) {\vector(-1,0){0}}
\put(190,120) {\vector(1,0){0}}
\put(170,160) {\vector(-1,0){0}}
\put(210,40) {\vector(-1,0){0}}
\put(230,80) {\vector(1,0){0}}
\put(210,120) {\vector(-1,0){0}}
\put(230,160) {\vector(1,0){0}}
\put(270,40) {\vector(1,0){0}}
\put(250,80) {\vector(-1,0){0}}
\put(270,120) {\vector(1,0){0}}
\put(250,160) {\vector(-1,0){0}}
\put(290,40) {\vector(-1,0){0}}
\put(310,80) {\vector(1,0){0}}
\put(290,120) {\vector(-1,0){0}}
\put(310,160) {\vector(1,0){0}}
\put(350,40) {\vector(1,0){0}}
\put(330,80) {\vector(-1,0){0}}
\put(350,120) {\vector(1,0){0}}
\put(330,160) {\vector(-1,0){0}}
\thinlines
\multiput(140,20)(-4,4){10}{\circle*{2}}
\multiput(100,60)(4,4){20}{\circle*{2}}
\multiput(180,140)(-4,4){10}{\circle*{2}}
\end{picture}
\label{fig:antiper}
\end{figure}

\vskip 10cm
\noindent
Figure 2: A typical lowest-energy state of the system
with $N$ even and anti-periodic boundary conditions.
The dotted line indicates the interface dividing the lattice into
two domains, each of which is an anti-ferroelectrically ordered
ground state.

\setcounter{figure}{2}
\begin{figure}[htb]
\epsfxsize = 11cm
\vbox{\vskip .8cm\hbox{\centerline{\epsffile{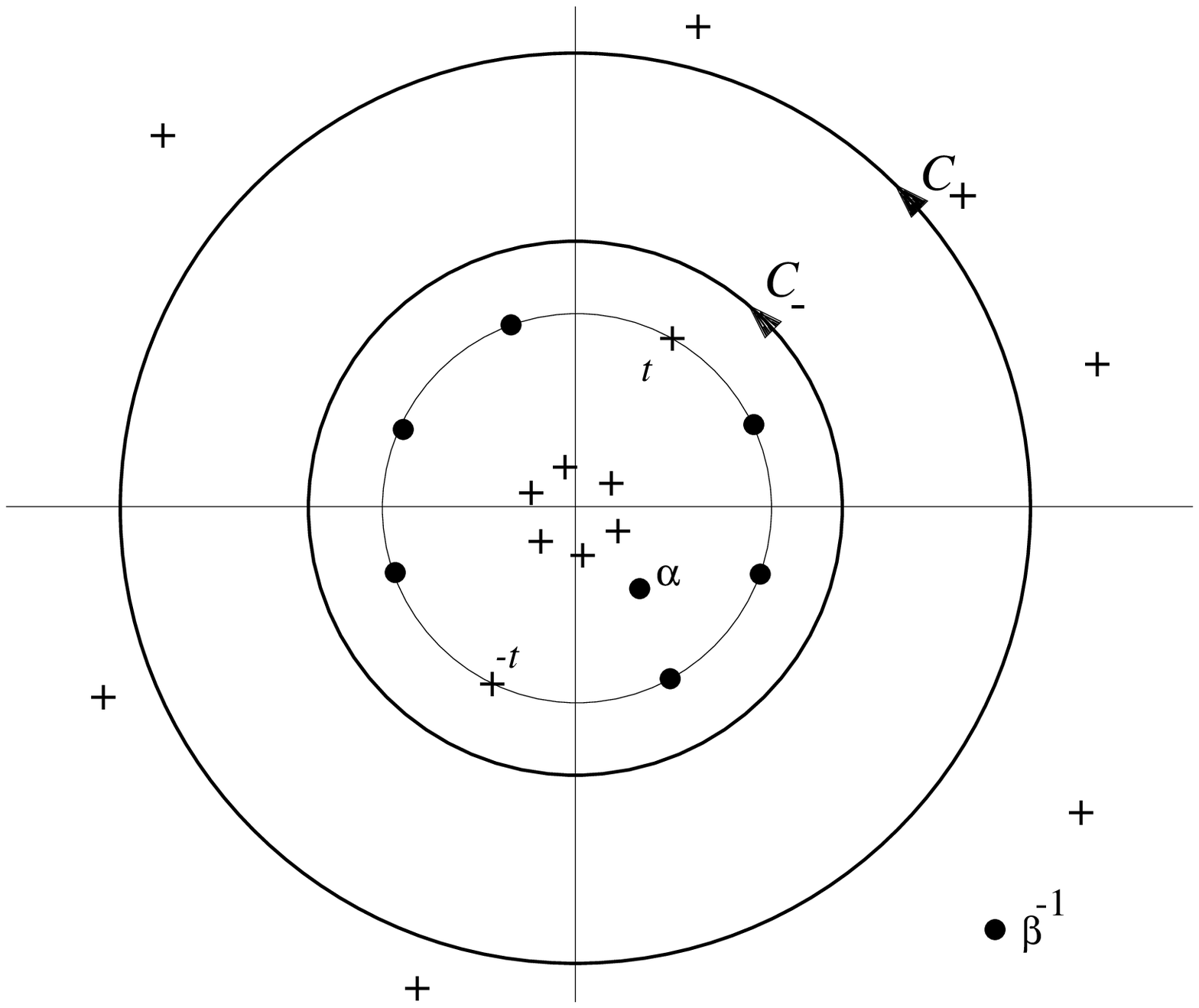}}}
%50%
\vskip .5cm \smallskip}
\caption{
The complex $z$-plane; the curves ${\cal
C}_{+}$ and ${\cal C}_{-}$ are indicated, with the unit circle lying
inside ${\cal C}_{-}$.
The zeros of $\tilde{Q}$ are indicated by $\bullet$ and the zeros of
$z^{-1}V(z)$ by $+$.
There are no zeros of either function in the annulus between the
contours ${\cal C}_+$ and ${\cal C}_-$.}
\end{figure}

\end{document}